\documentclass{article}

\usepackage{arxiv}

\usepackage[utf8]{inputenc} 
\usepackage[T1]{fontenc}    
\usepackage{hyperref}       
\usepackage{url}            
\usepackage{booktabs}       
\usepackage{amsfonts}       
\usepackage{amsmath}
\usepackage{nicefrac}       
\usepackage{microtype}      
\usepackage{graphicx}
\usepackage[numbers]{natbib}
\usepackage{doi}

\usepackage{float}

\title{Disease progression model anchored around clinical diagnosis in longitudinal cohorts: example of Alzheimer's disease and related dementia}

\date{}

\author{
    {\hspace{1mm}Jérémie~Lespinasse} \\
    Bordeaux Population Health research center \\
    Bordeaux School of Public Health \\
    CIC 1401-EC \\
    Univ. Bordeaux, Inserm \\
    Pôle de santé publique \\
    Centre Hospitalier Universitaire (CHU) de Bordeaux \\
	Bordeaux, France \\
	\texttt{jeremie.lespinasse@outlook.com} \\
	\And
	{\hspace{1mm}Carole~Dufouil} \\
	Bordeaux Population Health research center \\
    Bordeaux School of Public Health \\
    CIC 1401-EC \\
    Univ. Bordeaux, Inserm \\
    Pôle de santé publique \\
    Centre Hospitalier Universitaire (CHU) de Bordeaux \\
	Bordeaux, France \\
	\texttt{carole.dufouil@u-bordeaux.fr} \\
	\And
	{\hspace{1mm}Cécile~Proust-Lima} \\
	Bordeaux Population Health research center \\
    Bordeaux School of Public Health \\
    CIC 1401-EC \\
    Univ. Bordeaux, Inserm \\
	Bordeaux, France \\
	\texttt{cecile.proust-lima@u-bordeaux.fr} \\
}

\renewcommand{\headeright}{}
\renewcommand{\shorttitle}{Disease progression model anchored around clinical diagnosis in longitudinal cohorts}

\hypersetup{
pdftitle={A template for the arxiv style},
pdfsubject={stat.ME},
pdfauthor={Jeremie Lespinasse, Carole Dufouil, Cecile Proust-Lima},
pdfkeywords={multivariate mixed model, disease progression model, latent time, neurodegenerative disease, Alzheimer's disease},
}

\begin{document}

\maketitle

\begin{abstract}

\textbf{Background}
Alzheimer’s disease and related dementia (ADRD) are characterized by multiple and progressive anatomo-clinical changes including accumulation of abnormal proteins in the brain, brain atrophy and severe cognitive impairment. Understanding the sequence and timing of these changes is of primary importance to gain insight into the disease natural history and ultimately allow earlier diagnosis. Yet, modeling changes over disease course from cohort data is challenging as the usual timescales (time since inclusion, chronological age) are inappropriate and time-to-clinical diagnosis is available on small subsamples of participants with short follow-up durations prior to diagnosis. One solution to circumvent this challenge is to define the disease time as a latent variable.

\textbf{Methods}
We developed a multivariate mixed model approach that realigns individual trajectories into the latent disease time to describe disease progression. In contrast with the existing literature, our methodology exploits the clinical diagnosis information as a partially observed and approximate reference to guide the estimation of the latent disease time. The model estimation was carried out in the Bayesian Framework using Stan. We applied the methodology to the MEMENTO study, a French multicentric clinic-based cohort of 2186 participants with 5-year intensive follow-up. Repeated measures of 12 ADRD markers stemmed from cerebrospinal fluid (CSF), brain imaging and cognitive tests were analyzed.

\textbf{Results}
The estimated latent disease time spanned over twenty years before the clinical diagnosis. Considering the profile of a woman aged 70 with a high level of education and \textit{APOE4} carrier (the main genetic risk factor for ADRD), CSF markers of tau proteins accumulation preceded markers of brain atrophy by 5 years and cognitive decline by 10 years. However we observed that individual characteristics could substantially modify the sequence and timing of these changes, in particular for CSF level of A$\beta_{42}$.

\textbf{Conclusion}
By leveraging the available clinical diagnosis timing information, our disease progression model does not only realign trajectories into the most homogeneous way. It accounts for the inherent residual inter-individual variability in dementia progression to describe the long-term anatomo-clinical degradations according to the years preceding clinical diagnosis, and to provide clinically meaningful information on the sequence of events. 

\textbf{Trial registration}
clinicaltrials.gov, NCT01926249. Registered on 16 August 2013

\end{abstract}

\keywords{multivariate mixed model \and disease progression model \and latent time \and neurodegenerative disease \and Alzheimer's disease}

\section{Background}

Alzheimer’s disease and related dementia (ADRD) are characterized by progressive changes in multiple anatomo-clinical domains including decline in one or several cognitive functions (such as memory, language and executive function) leading to clinical dementia, functional dependency and death \cite{McKhann2011}. Alzheimer's disease (AD) neuropathology is identified by the abnormal accumulation of proteins that form amyloid plaques and tau neurofilaments in the brain \cite{Braak1991}. It is well established that brain vascular pathology contributes to cognitive impairment and dementia \cite{Zlokovic2020}. Especially small vessel disease (causing white matter lesions and silent brain infarcts) could double the risk of clinical dementia \cite{Azarpazhooh2018, Bos2018}. Progressive atrophy of some
brain regions, more specifically the hippocampus \cite{Schroder2016} or the medial temporal lobe \cite{Berron2020}, due to neurons deaths, was also highlighted to contribute to higher dementia risk. A decade ago, an hypothetical model of disease progression was proposed \cite{Jack2010} to temporally order these progressive changes. It postulates that amyloid and tau are involved in cellular mechanisms of protein deposition that would induce later neuronal dysfunction and brain structures atrophy. Decline in cognitive functions then appears as a result of the loss of neural tissues. Several studies found evidence supporting the initiating role of the amyloid protein on these pathological changes \cite{Jack2013, Bateman2012, Villemagne2013}, but these results were obtained among study participants at distinct clinical stages and no study provided a clear understanding of the anatomo-clinical changes over the entire natural history of the disease. In addition, the initial model completely ignored vascular contribution to cognitive impairment \cite{Jack2010}.

\begin{figure}[H]
    \includegraphics[width = 1\textwidth]{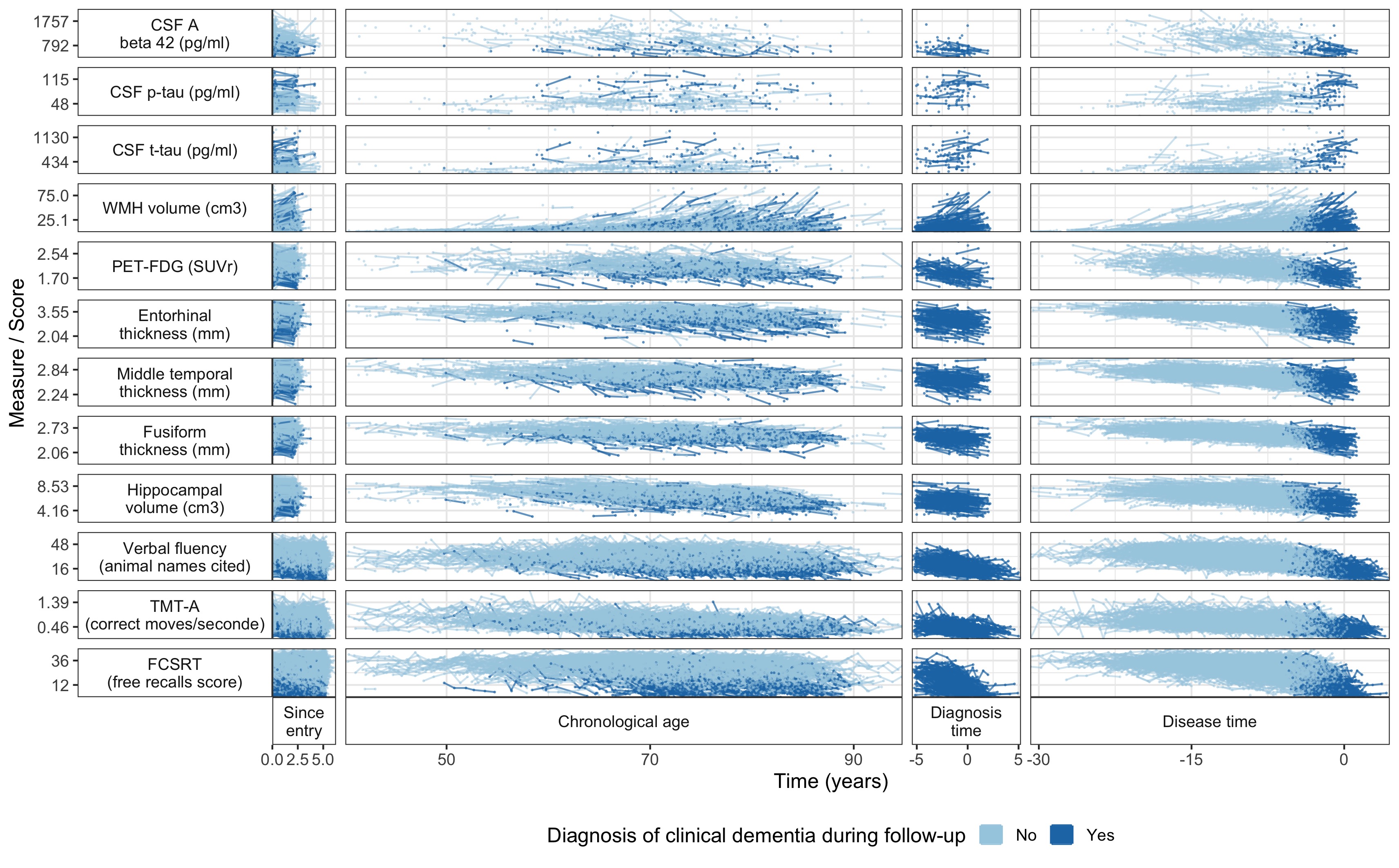}
    \caption{Individual trajectories of 12 markers in the MEMENTO Cohort, France, 2011-2019 (2186 participants, 286 incident cases of dementia) according to 4 different timescales: (A) time in the study (or follow-up time), (B) age, (C) time to diagnosis (available only for the incident cases) and (D) the latent disease time estimated using the disease progression anchored model.}
    \label{fig:intro}
\end{figure}

 Modeling disease progression from cohort studies is statistically straightforward in many diseases using the mixed model theory for instance \cite{Laird1982, Lindstrom1990}, but it faces a fundamental statistical challenge in ADRD. Indeed, since neuropathological changes likely occur 15 to 25 years before any clinical diagnosis can be reached \cite{Bateman2012, Vermunt2019}, confirming the sequence and timing of the associated neuropathological changes would require a follow-up of more than 20 years before diagnosis of clinical dementia which is only possible in population-based cohorts recruiting persons before middle age. Yet markers of neuropathological changes are mainly collected in clinical cohorts in which repeated measures of the most recent brain magnetic resonance imaging (MRI), brain positron emission tomography scanner (PET scan) and cerebrospinal fluid (CSF) derived biomarkers of ADRD before clinical diagnosis can be set up. This is the case with the MEMENTO cohort, a french nationwide clinic based study with 2323 participants followed up for 5 years, that gathers clinical examination, amyloid and tau biomarkers from cerebropsinal fluid, multiple brain images from MRI and PET scans (amyloid and glucose), and a neuropsychological tests battery. As illustrated in Figure \ref{fig:intro} (C) from the MEMENTO cohort data, modeling trajectories of the different markers according to the time to diagnosis in clinical ADRD cohorts usually both limits the analysis to the ultimate stages of the disease and reduces the sample size since most participants are not followed-up for more than 5 years (e.g. \cite{Petersen2010, Dufouil2017}). Alternative timescales, such as time since inclusion or chronological age (see individual markers trajectories in Figure \ref{fig:intro} (A) and (B), respectively), do not solve this temporal challenge for describing disease progression. Indeed, time since inclusion does not have any biological meaning, covers only a short period and is very heterogeneous because participants are included at different clinical stages. Although much more relevant in research on ADRDs and age-related disorders, chronological age still induces too much inter-individual heterogeneity as people do not age similarly and ADRD onset may arise at various ages.

In the absence of a relevant completely-observed timescale, latent disease progression models have been developed with the aim to directly retrieve the unobserved disease time from the data. These data-driven methods usually consist in re-aligning the participants trajectories according to the unobserved disease time by assuming that participants experience overall the same disease progression. After a first methodology proposed by Jedynak et al. \cite{Jedynak2012} to estimate a continuous disease time and describe the long-term progression of the biomarkers, many approaches have been developed and improved. Re-alignment of the individual trajectories into the disease progression time scale is managed using time-warping functions, through either an individual-specific time-shift \cite{Donohue2014, Li2017, Lorenzi2017, Raket2020, Garbarino2019, Kuhnel2021}, or an individual time-shift combined with individual rate of progression \cite{Jedynak2012, Bilgel2016, Marinescu2017, Schiratti2017, Koval2021} or the definition of an exponential progression score \cite{Bilgel2019}. Initially estimated individual by individual \cite{Jedynak2012}, most techniques now estimate the time-warping functions using random effects, thus entering the framework of nonlinear mixed models \cite{Li2017, Lorenzi2017, Raket2020, Garbarino2019, Kuhnel2021, Koval2021}. Beyond the use of time-warping functions, the different models also varied according to the type of information on participants considered (biological samples, brain images and/or neuropsychological evaluations) and according to the specification of the trajectories with sigmoid or exponential functions applied to Gaussian markers in their natural scales or after percentile transformations combined with linear mixed models. 

Despite the rise of disease progression models based on a latent disease time, none of the techniques directly considered the partially-observed information provided by the clinical dementia diagnosis. Yet, when based on a clinical expertise, diagnosis represents a landmark in the natural history of the disease that could help anchor the definition of the latent disease time along the actual ADRD process. We thus propose in this work a latent disease time model that directly incorporates the partially-observed diagnosis information to re-align the individual trajectories along the ADRD disease time. This Disease Progression Anchored Model (DPAM) is an extension of the latent time joint mixed effect model (LTJMM) developed by Li et al. (2017) \cite{Li2017} and estimated in the Bayesian framework using Stan. The methodology was applied to describe the progression of 12 biomarkers including markers of AD neuropathology, small vessel disease, brain atrophy and cognitive functioning in the French clinic-based MEMENTO study.

\section{Methods}

\subsection{The MEMENTO Cohort}

The MEMENTO cohort is a clinic-based study that recruited consecutively 2323 participants between April 2011 and June 2014 within the French national network of university-based memory clinics (Centres de Mémoires de Ressources et de Recherche [CMRR]). Participants were followed-up every 6 to 12 months during 5 years. Inclusion criteria required that participants were not demented, had a clinical dementia rating (CDR) $\leq$ 0.5, and performed 1 standard deviation worse than the subject’s own age, sex and educational-level group mean in one or more cognitive functions (from neuropsychological tests performed within 6 months preceding the screening phase). Participants with isolated subjective complaints were also eligible if aged 60 years or older. This study was performed in accordance with the Declaration of Helsinki. All participants provided written informed consent. The MEMENTO cohort protocol was approved by an ethics committee (“Comité de Protection des Personnes Sud-Ouest et Outre Mer III”; approval number 2010-A01394-35) and was registered in ClinicalTrials.gov (Identifier: NCT01926249).

The study protocol is described in details in \cite{Dufouil2017}. Every year, participants underwent a clinical evaluation that included an extensive battery of neuropsychological tests. Suspected cases of dementia during follow-up were reviewed by an independent expert committee and final clinical dementia diagnoses were established. At inclusion and at 24 months follow-up, all the patients were invited to undergo cerebral MRI, a 18F-fluorodeoxyglucose PET (FDG-PET) brain scan, and to have a lumbar puncture. The analytical sample consisted of the 2186 participants with at least one measure for one biomarker during the follow-up and without missing information on risk factors of interest in this work i.e. age, sex, education years and \textit{APOE} status (apolipoproteine E gene).

\subsection{Markers of ADRD}

Repeated measures of 12 markers of AD neuropathology, or small vessel disease, or brain atrophy or cognitive functioning were analyzed.

\subsubsection*{Biomarkers of AD neuropathology}

The three markers of AD neuropathology were the amyloid-$\beta$42 peptide (A$\beta$42), total tau (t-tau), and phosphorylated tau (p-tau181) measured from CSF using the standardized commercially available INNOTEST sandwich enzyme-linked immunosorbent assay (Fujirebio, Ghent, Belgium). 

\subsubsection*{Biomarkers of brain atrophy}

The markers of brain atrophy were cortical thickness in three regions associated with ADRD progression (middle temporal, enthorhinal, fusiform)\cite{Desikan2009} and hippocampal volume respectively measured from MRI T1-weighted with FreeSurfer\cite{Fischl2004} and SACHA\cite{Chupin2009}. Hippocampal volume was relative to the total intracranial volume. Glucose metabolism, a marker of neuronal loss, was measured by the mean FDG-PET uptake in AD-specific regions expressed as standard uptake value ratios (SUVr) \cite{Buchert2005, Habert2016}.

\subsubsection*{Biomarker of small vessel disease}

We used white matter lesions volumes as a marker of small vessel disease. MRI 2D-T2 FLAIR sequences analysis allowed to assess  white matter hyperintensities (WMH) volume using an automated and validated method \cite{Samaille2012}.

\subsubsection*{Cognitive Assessment}

Assessments results of three cognitive functions commonly impaired with the disease progression \cite{Weintraub2012} were included: 
\begin{itemize}
\item[-] episodic memory with the sum score of the 3 free recalls from the free and cued selective reminding test (FCSRT), a french adaptation of the Grober and Buschke test \cite{Grober1988}.
\item[-] semantic verbal fluency with the number of animals cited in 120 seconds \cite{Thurstone1987}. 
\item[-] executive functions with the number of correct moves per seconds at the trail making test A (TMT-A) \cite{Tombaugh2004}. 
\end{itemize}

\subsection{Disease progression anchored model}

Let's consider $K$ markers measured repeatedly over time. They are denoted $Y_{ijk}$ for the value of marker $k$ ($k = 1, ..., K$) for subject $i$ ($i = 1, ..., N$) at time $t_{ijk}$, with $j$ the occasion ($j = 1, ..., n_{ik}$). Time $t$ is the fully observed timescale: age or time since entry in the study in our case. The DPAM is defined in three steps: (i) define the latent disease time from the observed timescale, (ii) define a comparable scale for all the markers, (iii) define the multivariate mixed model for the marker trajectories according to the latent disease time. We describe each step below. 

\subsubsection*{Latent disease time definition}

A latent disease time can be generically defined as an individual-specific monotonic function $s_i(t)$ of the observed time $t$. In our framework, the latent disease time corresponds to the actual time since clinical diagnosis had it been made in continuous time. By denoting $T^*_i$ the actual unobserved time of clinical dementia, the latent disease time is defined as: 
\begin{equation}
s_i(t) = t - T^*_i
\end{equation}

This definition assumes there is no distortion of time between $t$ and $s$. The time in the disease $s$ is a shift of the observed time $t$ so that it is anchored to the actual time of clinical dementia: $s_i(T^* _ i )) = 0$. We assume that the actual time of clinical dementia is a latent variable with generic distribution $\mathcal{D}$. In the main analysis, we considered for instance a lognormal distribution: $\text{ln}(T^*_i) \sim {\mathcal{N}}(\mu_T ,\sigma_T ^{2})$. Without additional properties, the definition of this latent time shift is very standard and unrelated to the prior knowledge we may have about the time to clinical dementia. 

In cohorts that focus on ADRD risk factors and natural history, only a part of the participants is diagnosed with clinical dementia during the follow-up, either because some participants dropped out the study or died free of  clinical dementia diagnosis or were free of clinical dementia diagnosis by the end of the planned follow-up. However, whether a participant was diagnosed with clinical dementia or not, valuable information can be leveraged to anchor the latent time $T^* _ i$ to the actual time of clinical dementia. 

\subsubsection*{Prior knowledge on the diagnosis}

The clinical stage of the participants at the time of diagnosis is relatively homogeneous since clinical dementia is diagnosed by an independent committee of experts in the context of clinic based cohorts with intensive follow-up. Thus, time of clinical dementia diagnosis provides a reliable anchor time.

Let $D_i$ be the indicator that the participant had a confirmed diagnosis of clinical dementia during follow-up. For participants diagnosed with clinical dementia ($D_i = 1$) the observed diagnosis time $T^{\text{diag}}_i$ is likely in the neighbourhood of the actual unobserved time $T^* _ i$. For participants who dropped out free of clinical dementia ($D_i=0$), the time at the last clinical evaluation $T^{\text{last}}_i$ is likely to be smaller than the actual unobserved disease time $T^* _ i $. This was translated into the following constraints:
\vspace{1em}
\begin{itemize}
    \item[] $T^*_i > T^{\text{last}} - \epsilon_L$ for $D_i = 0$
    \item[] $T^{\text{diag}} - \epsilon_L ~<~ T^*_i ~<~ T^{\text{diag}} + \epsilon_U$ for $D_i = 1$
\end{itemize}
\vspace{1em}
where $\epsilon_L$ and $\epsilon_U$ are fixed scalars translating the lack of accuracy around the clinical evaluation and the diagnosis of clinical dementia. They have to be determined according to the study protocol and frequency of clinical evaluations. 

\subsubsection*{Severity scale and comparison of markers}

Independently from timescale definition, describing and comparing the sequence and speed of degradation across markers induces an additional challenge. Each marker has its own scale, and some, such as psychometric tests, are not necessarily Gaussian. Following previous works \cite{Donohue2014, Li2017, Lorenzi2017}, we transformed the raw markers data $Y_{ijk}$ into: 

(i) percentiles $P_{ijk}=F_{Y_k}(Y_{ijk})$ ($P_{ijk} \in [0,1]$) using the empirical cumulative distribution function $F_Y$ of each marker. This allowed to define a severity scale from 0 (minimum value) to 1 (maximum value) on which to compare the sequence of markers' impairments. Note that markers were flipped when necessary, so that higher values systematically indicated higher impairment (0 = best condition observed and 1 = worst condition observed). Since the empirical cumulative distribution function highly depends on sample characteristics, it may not translate equi-distributed levels of impairments. For instance, in the MEMENTO cohort, most patients remain at a very early clinical stage so that moderate to advanced stages would only correspond to the highest percentiles of the distribution. As suggested by other authors \cite{Li2017}, this was corrected by preliminarily applying individual weights ($w_i$) to the marker measures according to the clinical stage of the participant. This percentile transformation can be obtained by the weighted cumulative density function $P_{ijk}=F^w_{Y_k}(Y_{ijk},w_i)$ implemented in the Hmisc R package \cite{RHmisc}.

(ii) normalized values $\widetilde{Y}_{ijk}= \Phi^{-1}(P_{ijk})$ using the inverse of the Gaussian cumulative distribution function $\Phi^{-1}$. This allowed to apply multivariate linear mixed models for normal dependent variables $\widetilde{Y}$ and make predictions into the percentile scale using the inverse transformation $P_{ijk} =\Phi(\widetilde{Y}_{ijk})$. 

\subsubsection*{Multivariate linear mixed effects model}

We described the marker trajectories in the normalized scale $\widetilde{Y}_{ijk}$ according to the latent disease time $s_i(t_{ijk})$ using the following multivariate linear mixed model:
\begin{equation}
\widetilde{Y}_{ijk} = \boldsymbol{F}(s_i(t_{ijk}))^\top \boldsymbol{\beta}_k + \boldsymbol{X}_{i}(t_{ijk}) \boldsymbol{\gamma}_k + \boldsymbol{F}(s_i(t_{ijk}))^\top \boldsymbol{u_{ik}} + \varepsilon_{ijk}
\label{MLMM}
\end{equation}

where $\boldsymbol{F}$ is a basis of time functions defining the shape of the trajectory according to the disease time. Associated with $\boldsymbol{\beta}_{k}$, it gives the mean trajectory of normalized marker $k$ (for the reference profile of covariates). $\boldsymbol{X}_{i}(t_{ijk})$ are adjustment covariates associated with fixed effects $\boldsymbol{\gamma}_k$ and $\varepsilon_{ijk}$ are the independent Gaussian error of measurement with marker-specific variance $\sigma_{\varepsilon_k}^2$. Finally $\boldsymbol{u}_{ik}$ are the individual-and-marker-specific random effects defining the individual departure from the marker-mean trajectory. We assumed $\boldsymbol{u}_{ik} \sim \mathcal{N}(0,\boldsymbol{B}_k)$ with $\boldsymbol{B}_k$ an unstructured variance covariance matrix. Random effects and errors are assumed independent. In addition, we assumed that the markers-specific random deviations were independent across markers so that the latent time-shift captured the inter-markers correlation.

\subsection*{DPAM specification for the MEMENTO Cohort}

In the application to the MEMENTO Cohort, we considered a linear marker-specific trajectory ($\boldsymbol{F}(s) = (1, s)^\top$) and constrained $\boldsymbol{\beta}_k \geq 0$ to impose a mean degradation over time for all the biomarkers. We also included a random slope only for the neuropsychological tests. This was to prevent any numerical non-identifiability issues for MRI and CSF markers where a maximum of two measures was collected. In addition, we considered as adjustment covariates: age, sex, years of education and \textit{APOE4} status. Finally, we added an indicator of first visit for the neuropsychological tests to correct for the first passing effect \cite{Vivot2016}.

Given clinical dementia diagnoses were performed every 6 months in the cohort, we set the lack of accuracy around clinical evaluation to $\epsilon_L = \epsilon_U = 1.5$ years in the main analysis. 

The individual weights used to define the percentiles of the marker severity scales were derived from four clinical stages according to the CDR-SB (CDR sum of the boxes) score at entry in the study: CDR-SB = 0, N = 784, $w_i$ = 2.80; CDR-SB = 0.5, N = 794, $w_i$ = 2.74; CDR-SB = 1, N = 323, $w_i$ = 6.81; and CDR-SB $>$ 1, N = 285, $w_i$ = 7.57.

\subsection{Estimation procedure}

The estimation of our disease progression model was done in the Bayesian framework using Hamiltonian Monte Carlo  No-U-turn sampling algorithm (HMC-NUTS) \cite{Hoffman2014} to approximate the posterior distribution of the parameters with Markov Chain Monte Carlo (MCMC). We used Stan software (version 2.20.0) \cite{carpenter2017stan, stanref} through the CmdStan interface with parallel computations on both the chains and the individuals. A commented version of our program, freely adapted from LTJMM \cite{Li2017}, is available at \texttt{https://github.com/jrmie/dpm\_anchored}.

\subsubsection*{Prior distributions}

We considered standard weakly-informative priors for the multivariate mixed model parameters in equation \eqref{MLMM} with for all $k=1,...,K$: each element of $\boldsymbol{\beta}_k$ and $\boldsymbol{\gamma}_k$ following $\mathcal{N}(0,10^2)$ (with $\boldsymbol{\beta}_k$ imposed to be positive), and $\sigma_{\varepsilon_k}$ and the variances of the random-effects $\boldsymbol{u}_{ik}$ following $\text{half-Cauchy}(0,2.5)$.
For the latent disease time, we assumed the following distribution to incorporate the $\epsilon_L$ constraint and allow for negative $T^*_i$: $\text{ln}(T^*_i + \epsilon_L) \sim \mathcal{N}(\mu_{T\epsilon},\sigma_{T\epsilon})$ with $\mu_{T\epsilon} \sim \mathcal{N}(10,10^2)$, and $\sigma_{T\epsilon} \sim \text{half-Cauchy}(0,2.5)$.

\subsubsection*{Posterior summaries}

We ran 4 chains of 6000 iterations burn-in and 2000 iterations for sampling, and we retained 1 iteration every 4 iterations to avoid auto-correlation issues from consecutive samples. Thus we approximated the posterior distribution with $D$=2000 iterations (500 by chain) and reported posterior means and 95\% confidence intervals (95\%CI) of the parameters.

\subsubsection*{Diagnostic checks}

Diagnostic tools of Stan were used to evaluate the estimation procedure: convergence of the MCMC with the Gelman and Rubin \cite{Gelman1992} potential scale reduction statistic $\hat{R}$ which compares variances between and within chains and effective sample size ratio ($ESS$) \cite{Gelman2013} which estimates sample size without any auto-correlation. These indicators were considered as satisfied if $\hat{R} <1.05$ and $ESS/D \geq 0.1$) for all parameters of the model.

\subsubsection*{Sensitivity analyses}

We assessed the influence of our definition of the latent disease time and the associated constraints in sensitivity analyses. Specifically, we compared our DPAM using the actual dementia time defined according to a lognormal distribution and the use of prior information on the observed diagnoses times with the non-anchored disease progression model specification in which $T_i^*$ followed a Gaussian distribution without any constraint. We also evaluated the stability of the results when considering weaker constraints ($\epsilon_L =\epsilon_U=3$  year) to guide the estimation of the disease time. The comparison was based on the residual root mean squared error (RMSE).

\subsection{Predictions of the biomarkers' mean trajectories in the severity scale}

A central output of this methodology is the description of biomarkers mean trajectories according to latent disease time $s$.  Let define $ P_{ik} (s_i(t_{ijk})) = P_{ijk} $ and $ \widetilde{Y}_{ik} (s_i(t_{ijk})) = \widetilde{Y}_{ijk}$. The mean trajectory of biomarker $k$ for a covariate profile $\boldsymbol{x}$ (independent of time for simplicity) according to latent disease time $s$ in the severity scale is:
\begin{equation}
    \mathbb{E}(P_{ik} (s) |_{X_i(t)=\boldsymbol{x}}) = \int \Phi(\widetilde{y}) f_{\widetilde{Y}(s)|_{X_i(t)=\boldsymbol{x}}} (\widetilde{y})d\widetilde{y}
\end{equation}
where $f_{\widetilde{Y}(s)|_{X_i(t)=\boldsymbol{x}}}$ is the density function of $\widetilde{Y}(s)$ given $X_i(t)=\boldsymbol{x}$. At each iteration $d$ of the MCMC, this integral can be approximated by the Monte Carlo technique as $\mathbb{E}(P_{ik} (s) |_{X_i(t)=\boldsymbol{x}}) \approx \hat{P}_{ik}(s,\boldsymbol{x}) = \dfrac{1}{M} \sum_{m=1}^M  \Phi(\widetilde{y}_m)$ where $\widetilde{y}_m$ is randomly drawn from $\mathcal{N} \left (\boldsymbol{F}(s) \beta_k^{(d)} + \boldsymbol{x} \boldsymbol{\gamma}_k^{(d)}, \boldsymbol{F}(s)^\top B_k^{(d)}\boldsymbol{F}(s)^\top + \sigma_{\varepsilon_k^{(d)}}^2 \right)$ where $^{(d)}$ indicates the value of the parameters at the MCMC iteration $d$. We considered $M=1000$.

The mean and its 95\%CI over the iterations are retained to describe the mean biomarker trajectory $\hat{P}_ik(s,\boldsymbol{x})$ over latent disease time for covariate profile $\boldsymbol{x}$ between $s=-30$ years before dementia to $s=5$ years after.

\section{Results}

Participants in the analytical sample (N=2186) were aged 70.9 years (SD=8.7) on average, 61.7\% were women, 39.2\% had more than 12 years education and 29.9\% carried at least 1 allele $\epsilon$4 of \textit{APOE} gene (\textit{APOE4}, a major genetic risk factor for ADRD). Additional description of the analytical sample at inclusion is reported in Table 1. During the 5-year follow-up, 284 participants developed dementia.

A description of the distribution markers at baseline is reported in Table \ref{Tab1} and individual trajectories are displayed in Figure \ref{fig:intro} (A,B,C). Almost all participants had a least one cognitive measure at baseline, 2047 participants had volumetric MRI, 1236 had PET-FDG and 342 had CSF biomarkers. The number of repeated measures varied between 0 and 2 for CSF, MRI and FDG-PET, 0 and 6 for cognitive measures.

\subsection{Estimated latent disease time}

\begin{figure}[H]
    \includegraphics[width=1\textwidth]{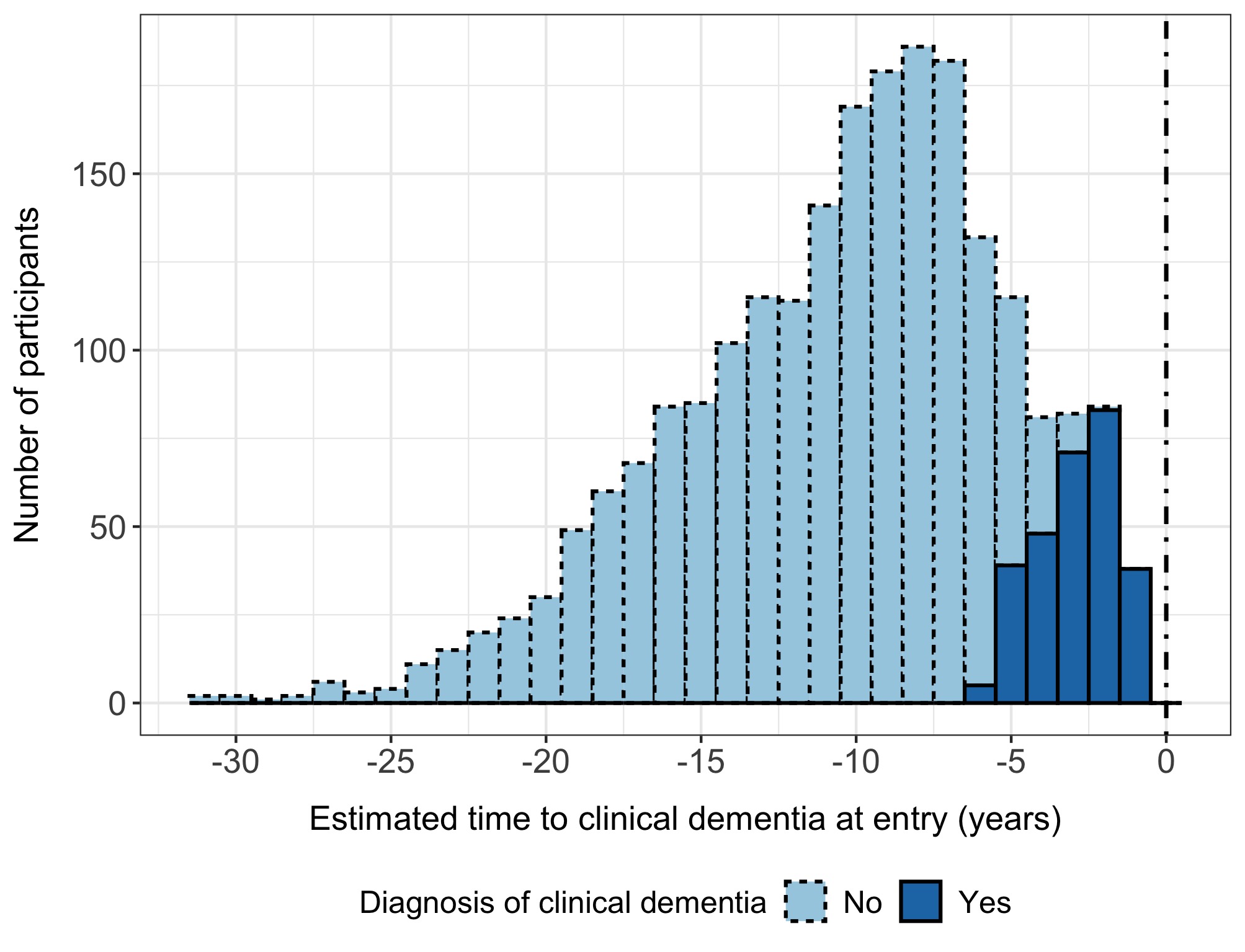}
    \caption{Posterior distribution of the estimated individual time to actual dementia at entry in the MEMENTO Cohort (France, 2011-2019, N=2186) according to the last dementia diagnosis status.}
    \label{fig:tdem}
\end{figure}

Individual estimated times of actual clinical dementia $T_i^*$ were used to display the estimated delay to actual clinical dementia from entry in the cohort $s_i(0)$ (Figure \ref{fig:tdem}). For instance, $s_i(0) = -3$ corresponds to an estimated actual time of clinical dementia of 3 years after entering the cohort.

Figure \ref{fig:intro} (D) displays the individual observed markers' trajectories according to the estimated clinical dementia time. Participants entered the cohort on average 10.3 years before the estimated actual clinical dementia onset with a range from 0.74 to 30.8 years. Among incident dementia cases, time to dementia onset varied between 0.74 and 6.15 years. These times to dementia were very close to the observed clinical dementia diagnoses with an inaccuracy ranging from 1.01 years prior to the estimated time and 0.59 years after the estimated time (while the constraints allowed up to +/- 1.5 years).

\subsection{Covariates association with the biomarker levels}

\begin{figure}[H]
    \includegraphics[width=1\textwidth]{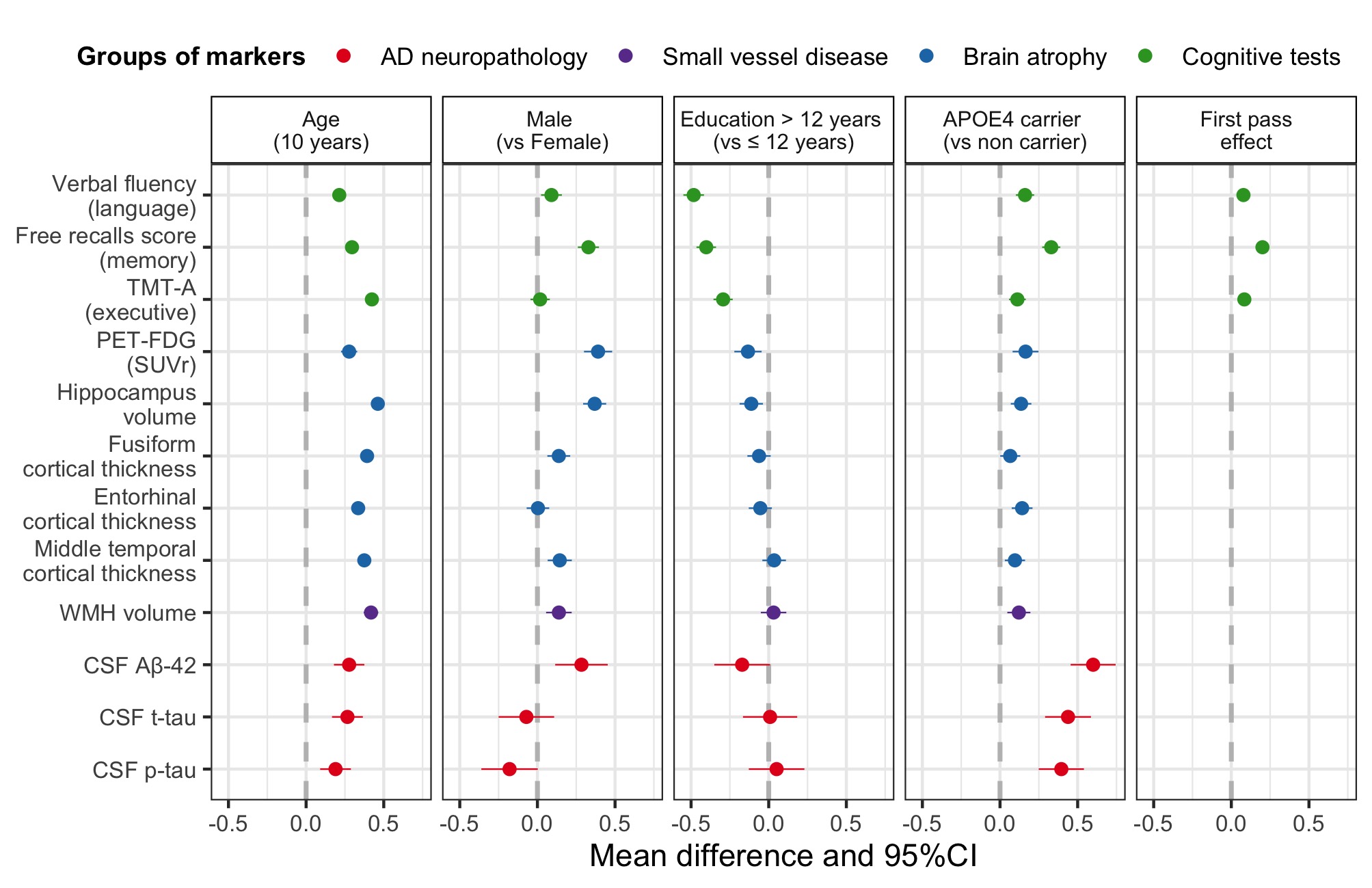}
    \caption{Estimated association of age, sex, education, \textit{APOE4} status, and first practice effect with each of the 12 biomarkers in the normalized scale, the MEMENTO Cohort, France, 2011-2019 (N=2186).}
    \label{fig:covar}
\end{figure}

Figure \ref{fig:covar} displays the mean and corresponding 95\%CI of the covariates associations with each biomarker. All coefficients $\gamma_k$ are reported in standard deviation (SD) of the considered marker and adjusted for the other covariates. Age was significantly associated with worse levels for all markers. 

The association with sex showed a substantially greater severity for men in memory domain (mean difference (MD)=0.33, 95\%CI=[0.26, 0.40]), hippocampal volume (MD=0.37, 95\%CI=[0.29, 0.44]) and FDG-PET (MD = 0.39, 95\%CI = [0.30, 0.48]), and to a lesser extent on amyloid level (CSF A$\beta_{42}$) (MD=0.28, 95\%CI=[0.12, 0.45]), WMH volume (MD=0.14, 95\%CI=[0.05, 0.22]) and cortical thicknesses of fusiform (MD=0.14, 95\%CI=[0.06, 0.21]) and middle temporal (MD=0.14, 95\%CI = [0.07, 0.22]). Men tended to have higher level of p-tau (MD=-0.18, 95\%CI=[-0.36, 0.00]). 

High education ($>$12 years) was related to substantially better scores at cognitive tests: memory (MD=-0.40, 95\%CI=[-0.46, -0.434]), language (MD=-0.48, 95\%CI=[-0.55, -0.42]) and executive function (MD=-0.29, 95\%CI=[-0.36, -0.23]); high education was also slightly associated with lower degradation in FDG-PET (MD=-0.13, 95\%CI=[-0.22, -0.05]) and hippocampal volume (MD=-0.11, 95\%CI=[-0.19, -0.04]) but it was not related to cortical thicknesses, WMH volume or CSF markers.

\textit{APOE4} carriers displayed on average worse results on ADRD biomarkers with larger differences for A$\beta_{42}$ (MD=0.60, 95\%CI=[0.45, 0.74]), p-tau (MD=0.39, 95\%CI=[0.25, 0.54]) and t-tau (MD=0.44, 95\%CI=[0.29, 0.59]) in CSF.

\subsection{Trajectories of the markers in the latent disease time}

\begin{figure}[H]
    \includegraphics[width=1\textwidth]{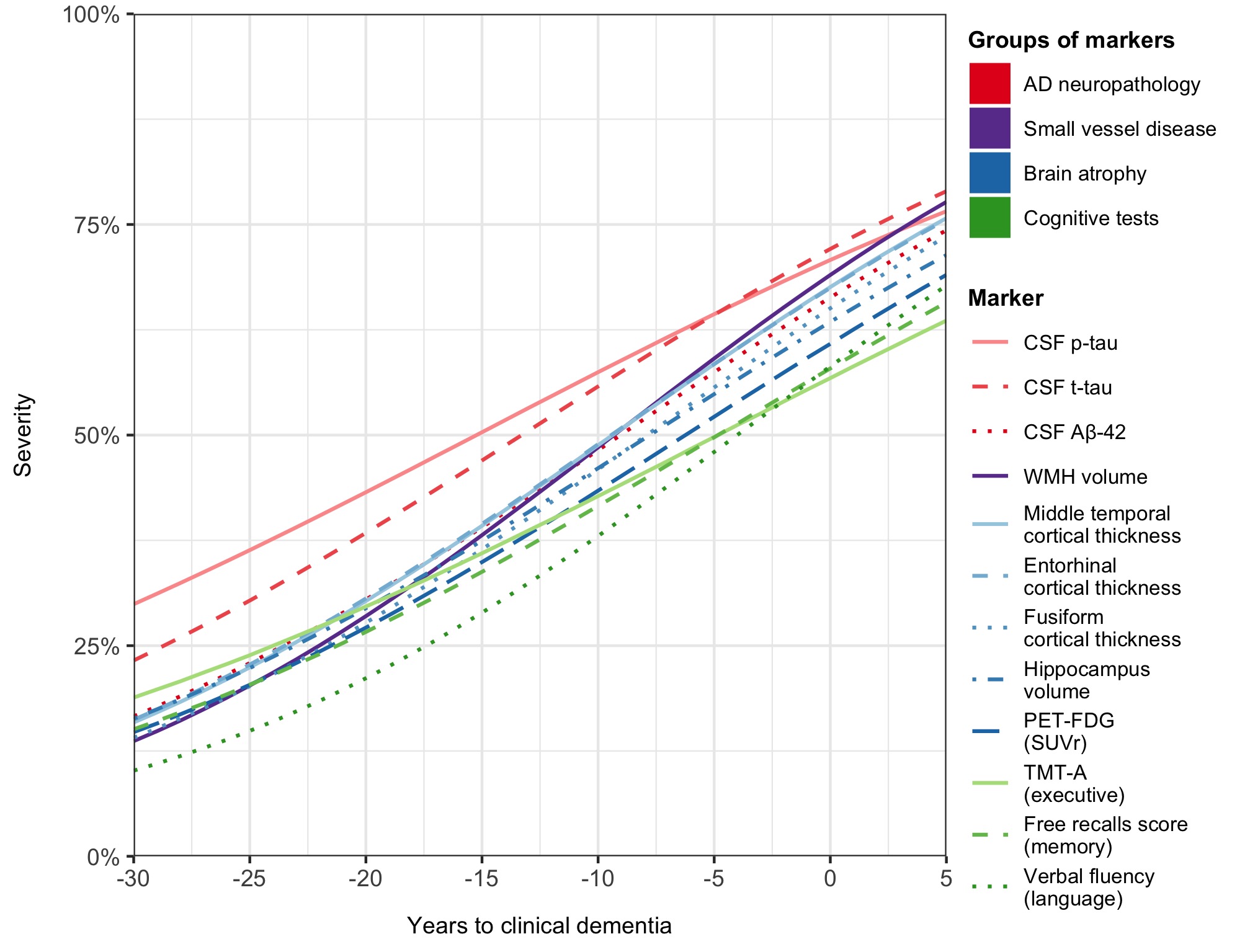}
    \caption{Mean trajectories of the 12 biomarkers of progression in the percentile scale according to latent disease time for a women of 70 years old, with more than 12 years of education and \textit{APOE4} carrier, the MEMENTO Cohort, France, 2011-2019  (N=2186).}
    \label{fig:pred}
\end{figure}

Figure \ref{fig:pred} displays the averaged trajectories of markers' progressions, between 30 years prior to clinical dementia to 5 years after, for a typical participant: woman of 70 years old, with more than 12 years of education and carrier of \textit{APOE4} allele. For better clarity 95\%CI are not shown.

\subsubsection*{Marker severities}

Thirty years prior to clinical dementia, all the markers were on average at low levels of severity (below 25\%) with the highest levels for total tau and p-tau (23\% (95\%CI=[0.14,0.32]) and 30\% (95\%CI=[0.20,0.40]), respectively). In comparison, the average 30\% severity level was reached for A$\beta_{42}$, WMH volume, volumetric neuroimaging (cortical thicknesses, hippocampal volume, FDG-PET), memory and executive functioning about 10 to 12 years later (that is 18-20 years prior to clinical dementia) and about 15 years later for verbal fluency. Total tau and p-tau remained the more impaired markers at all times although they degraded more slowly. WMH volume showed the fastest degradations with cortical thicknesses.

\subsubsection*{Order of marker changes}

\begin{figure}[H]
    \includegraphics[width=1\textwidth]{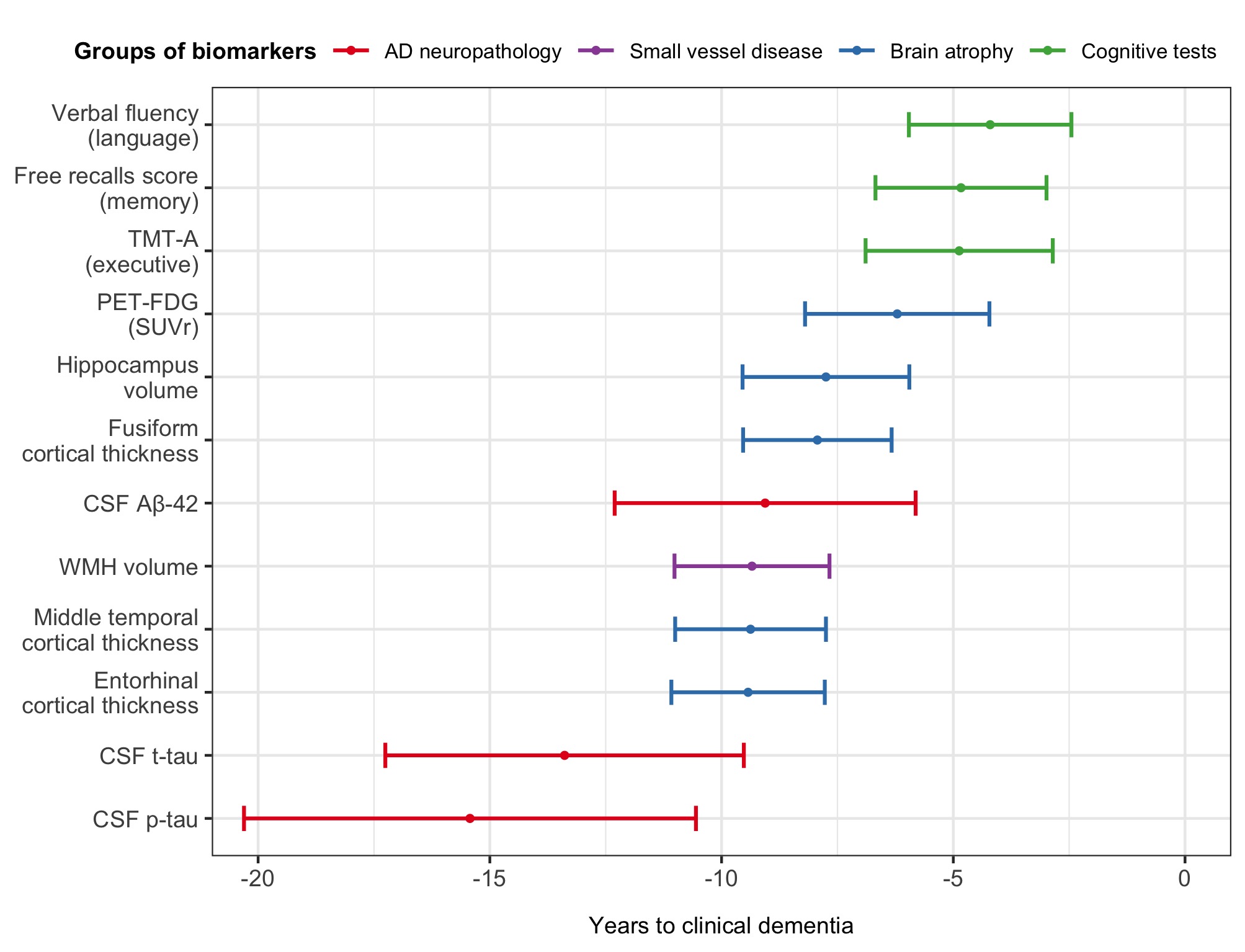}
    \caption{Ordering sequence and uncertainty (95\%CI) of the biomarkers reaching the moderate severity for a women of 70 years old, with more than 12 years of education and \textit{APOE4} carrier, the MEMENTO Cohort, France, 2011-2019 (N=2186).}
    \label{fig:pred_seq}
\end{figure}

Figure \ref{fig:pred_seq} summarizes the order in which the markers reach 50\% severity with the corresponding uncertainty (95\%CI) in the time scale of the disease (years before clinical dementia) for the same typical covariate profile as previously. According to the weighted severity scale the 50th percentile may give an indication of the entry into moderate severity. This level was first reached by p-tau and total-tau 15.4 (95\%CI=[10.6, 20.3]) and 13.4 (95\%CI=[9.5,17.3]) years before clinical dementia. About 5 years later the moderate severity was reached by A$\beta$42 along with WMH volume, and cortical thicknesses of the middle temporal and entorhinal regions: respectively 9.1 (95\%CI=[5.8, 12.3]), 9.3 (95\%CI=[7.7, 11.0]), and 9.4 (95\%CI=[7.8, 11.1]) years before clinical dementia. Then fusiform cortical thickness, hippocampus atrophy and glucose metabolism followed 1.5 to 3 years later with moderate severity reached 7.9 (95\%CI=[6.3,9.5]), 7.8 (95\%CI=[5.9,9.5]) and 6.2 (95\%CI=[4.2,8.2]) years before clinical dementia, respectively. Finally, cognitive tests reached the moderate severity about 10 years after the p-tau in CSF. That was 4.2 (95\%CI=[2.5,6.0]), 4.8 (95\%CI=[3.0,6.7]) and 4.9 (95\%CI=[2.9,6.9]) years before clinical dementia for language, memory and executive function, respectively.

\subsubsection*{Covariate profiles}

\begin{figure}[H]
    \includegraphics[width=1\textwidth]{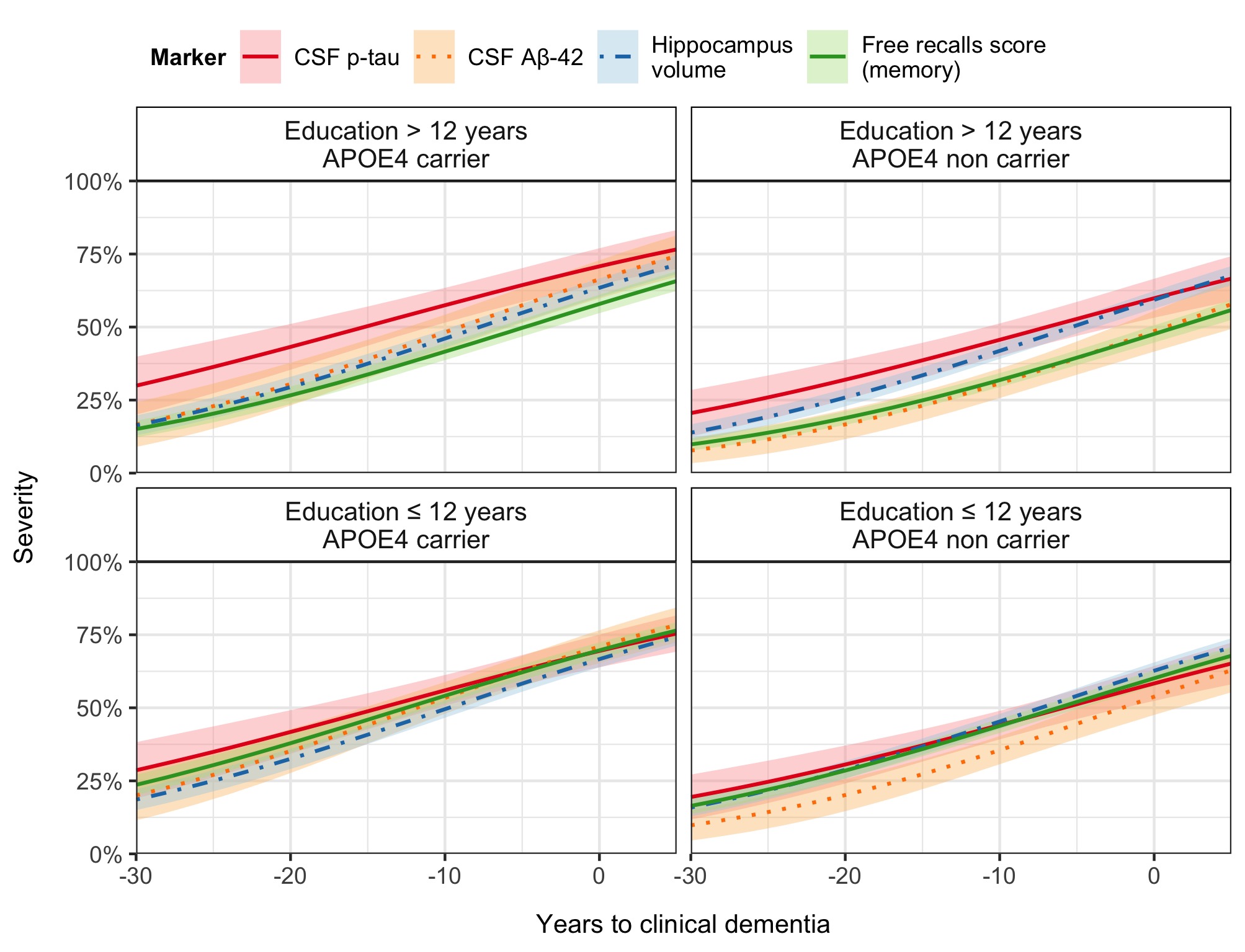}
    \caption{Average trajectories of 4 markers progression (A$\beta$42, p-Tau, hippocampal volume and FCSRT) according to latent disease time in the percentile scale for the 4 covariate profiles (education and \textit{APOE4}), the MEMENTO Cohort, France, 2011-2019 (N=2186).}
    \label{fig:pred_x}
\end{figure}

Because of the differential effect of the covariates on the markers, the degradation sequence differed according to the profile participants. To give a better sense of the heterogeneity of the sequence, we displayed in Figure \ref{fig:pred_x} the averaged trajectories (along with 95\%CI) of 4 landmark biomarkers (p-tau level, A$\beta_{42}$ level, hippocampal volume and memory test score) according to education years and \textit{APOE4} status. 

The anteriority of p-tau degradation was found mainly among the high education groups. Memory impairment progressed years later (among highly educated profiles) or contemporaneously (among less educated profiles) with p-tau level and hippocampus atrophy. A$\beta$42 level was the most variable marker in the sequence. It reached moderate severity level years later hippocampus atrophy and even after memory impairment in the profile \textit{APOE4} non carriers and low education while the degradation of A$\beta$42 marker was at about the same time as the one of hippocampus atrophy for \textit{APOE4} carriers. 

\subsection{Sensitivity analyses}

The fit to the data was unchanged when considering a larger uncertainty around the observed clinical diagnoses with $\epsilon_L$=$\epsilon_U$=3 years (RMSE=0.0858 for both models, see e-Figure 1 in supplementary materials for RMSE per marker), and the results remained virtually the same.
We also compared our DPAM that assumed a log-normal distribution for the latent disease time and anchored the latent disease time along the clinical dementia diagnosis with a non-anchored disease progression model (similar to the LTJMM methodology \cite{Li2017}) in which the latent disease time definition was completely data-driven and the latent time shift distribution was assumed as normal. In this non-anchored model, the latent disease time is centered on the average stage of the analytical sample at entry in the cohort. The non-anchored approach performed very similarly as our DPAM with RMSE = 0.0825 and RMSE = 0.0858, respectively. The slight gain in RMSE of the non-anchored model was due to a slightly better fit of the neuropsychological data  (see e-Figure 1 in the supplementary materials for a RMSE separated by marker). In this model, the estimated latent disease times of individuals diagnosed with clinical dementia were very far from the actual time to clinical diagnosis with a span over 15 years (e-Figure 2 in the supplementary materials). Indeed, as completely data-driven, the latent time shift was determined as the one homogenizing at most the data, and it was more influenced by the neuropsychological markers than MRI and CSF markers as the former brought much more information with more repeated measures. This underlines the importance of anchoring the model to realign the trajectories in link with the patient staging rather than only the inter-marker correlation.

\section{Discussion}
 
We developed a disease progression model to describe the markers' trajectories of the anatomo-clinical dimensions identified in ADRD progression towards clinical dementia. Using the intensive follow-up data of the French MEMENTO Cohort, we identified a large variability in the patients staging at study entry with an estimated time to actual clinical dementia spanning over 30 years. The sequence of markers progression substantially varied according to education and \textit{APOE4} status. However, we consistently identified p-tau as the first marker showing a pathological progression years before the onset of structural damage visible at brain imaging. Moreover, white matter hyperintensities, occurring concomitantly to regional brain atrophies, seemed to progress faster than other markers. 

Compared to the rich literature on disease progression modelling using latent disease times (\cite{Jedynak2012, Lorenzi2017, Li2017, Bilgel2019, Raket2020}), our adopted approach goes one step further. As previous works, we defined the latent disease time as an individual latent time shift shared by the disease markers and estimated it from the data. However, we also leveraged prior information on the clinical diagnoses to guide the estimation of the latent time-shift towards an actual clinical dementia diagnosis. As shown in the sensitivity analyses, without this prior information, the latent time-shift may over homogenize the trajectories of the markers. In contrast, anchoring the definition of latent time shift around the observed clinical diagnosis made it possible to realign markers trajectories around the actual time of clinical dementia (which is an important step in the clinical management of patients) while preserving the inherent heterogeneity in disease progression across individuals. In contrast with the literature, we also chose a lognormal distribution for the latent time shift (rather than a normal distribution) to allow for a longer tail in the time shift distribution, in agreement with real life data. This is in line with time-to-event analyses where lognormal distribution is usually preferred over normal distribution \cite{JGadda2006}.                         

In the MEMENTO cohort, estimates of individual disease times to clinical dementia at study entry extended over decades, a consistent result with previous studies \cite{Vermunt2019}. Sequence and timing of markers along the natural history of the disease were partially consistent with the theoretical model of Jack et al. \cite{Jack2010} and we found major differences in the sequence and timing of the markers according to the individual characteristics. Our findings supported that CSF p-tau showed increasing severity years before the degradation of glucose metabolism and brain atrophy on neuroimaging. Structural brain changes also preceded worsening of cognitive function. While amyloid deposit is widely considered as the initial cause of Alzheimer's disease \cite{Jack2018}, timing of CSF A$\beta$42 was unclear as moderate severity was reached later than for CSF p-tau, a consistent result with previous disease progression model \cite{Li2018}. Indeed, timing of amyloid degradation substantially varied according to \textit{APOE4} status, thus contributing to the discussion challenging the central role of amyloid peptide in the natural history of the disease \cite{Frisoni2022}. Our results also reinforce the hypothesis of small vessel disease contribution to cognitive impairment and dementia \cite{Zlokovic2020}, as volume of white matter hyperintensities is the most rapidly deteriorating marker and contemporaneous with the degradation of cortical thicknesses years before cognitive impairment.

As any disease progression model, our approach relies on parametric assumptions. First, we considered a linear trajectory after a transformation of the markers into normal framework in agreement with previous works \cite{Li2017}. Alternative sigmoid models usually model a generalized logistic trajectory \cite{Jedynak2012} directly in the severity scale. Second, to distinguish the inter-marker correlation due to the disease staging from the intra-marker correlation, we assumed that the latent time shift captured all the correlation shared across markers and considered that marker-specific random effects were independent between markers. This assumption could be relaxed by allowing some correlation between subsets of markers, for instance MRI-derived markers or neurospychological tests.  Finally, we accounted for differences across covariate profiles through a global effect on each marker's severity. Although of interest, considering interactions between individual characteristics and rate of marker degradation would substantially complicate the model and the estimation procedure due to the higher number of additional parameters to estimate. A few progression models considered a covariate effect on the disease time \cite{Raket2020, Kuhnel2021} rather than on each marker measure separately. This may be interesting for exploring covariates that may delay the progression towards dementia. However, as found in our application, some covariates may differentially modulate markers’ trajectories and the sequence of markers’ degradation. This was the case for years of education that showed large differences only in the neuropsychological tests, as an illustration of the concept of cognitive reserve \cite{Arezana2018}.

Finally, by using the mixed model theory, our results are robust to missing at random (MAR) assumption for missing data which is plausible.  

\section{Conclusion}

Disease progression models  allow for the characterization of the complete natural history of a disease when observed time is not relevant as it is the case for ADRD. Applied on the MEMENTO clinic-based cohort, we showed that the shift of individual trajectories into a latent disease time scale extended over 30 years prior to dementia clinical diagnosis. This original work brings new insights in the understanding of the natural history of ADRD biomarkers both as we used information from actual diagnosis time of clinical dementia to estimate the latent time underlying the long term progression of the markers and as we based our work on a large cohort whereas most published work rely on ADNI data. 

\section*{Appendix}

\subsection*{Funding}
The MEMENTO cohort is funded by the Fondation Plan Alzheimer (Alzheimer Plan 2008‐2012), and the French Ministry of Research (MESRI, DGRI) through the Plan Maladies Neurodégénératives (2014‐2019). This work was also supported by CIC 1401‐EC, Bordeaux University Hospital (CHU Bordeaux, sponsor of the cohort), Inserm, and the University of Bordeaux. This work received funding from the French National Research Agency (ANR) as part of the Investment for the Future Programme ANR-18-RHUS-0002. The MEMENTO cohort has received funding support from AVID, GE Healthcare, and FUJIREBIO through private–public partnerships. The Insight-PreAD substudy was promoted by INSERM in collaboration with the Institut du Cerveau et de la Moelle Epinière, Institut Hospitalo-Universitaire, and Pfizer and has received support within the “Investissement d’Avenir” (ANR-10-AIHU-06) program. The funders had no role in study design, in data collection, analysis, and interpretation, or in writing of report. The corresponding author had full access to all the data in the study and had final responsibility for the decision to submit for publication.

\subsection*{Abbreviations}
A$\beta_{42}$ = amyloid beta 42, ADRD = Alzheimer's disease and related dementia, \textit{APOE4} = allele $\epsilon$4 of the apolipoproteine E gene, CDR = clinical dementia rate, CDR-SB = clinical dementia rate sum of the boxes, CI = confidence interval, CSF = cerebrospinal fluid, DPAM = disease progression anchored model, FCSRT = free and cued reminding test, FDG = fluorodeoxyglucose, LTJMM = latent time joint mixed model, MD = mean difference, MRI = magnetic resonance imaging, PET = positron emission tomography, p-Tau = phosphorylated tau, RMSE = root mean squared error, SUVr = standard uptake value ratio, TMT-A = trail making test A, t-Tau = total tau, WMH = white matter hyperintensities.

\subsection*{Availability of data and materials}
MEMENTO data access request is available via the Dementia Platform UK Data Access application form (\texttt{https://portal.dementiasplatform.uk/Apply}) or via the MEMENTO Secretariat (\texttt{sophie.lamarque@u-bordeaux.fr}). The Stan script used for the model specification are available at \texttt{https://github.com/jrmie/dpm\_anchored}.

\subsection*{Ethics approval and consent to participate}
This study was performed in accordance with the guidelines of the Declaration of Helsinki. The MEMENTO study protocol has been approved by the local ethics committee ("Comité de Protection des Personnes Sud-Ouest et Outre Mer III"; approval number 2010-A01394-35). All participants provided written informed consent.

\subsection*{Competing interests}
The authors declare that they have no competing interests.

\subsection*{Authors' contributions}
JL, CD, and CPL conceived and designed the study; CD collected the data; JL and CPL developed the statistical methodology; JL, CD, and CPL analyzed the data and wrote the paper. JL had full access to all of the data in the study. All authors take responsibility for the integrity of the data and the accuracy of the data analysis. All authors read and approved the final manuscript.

\bibliographystyle{vancouver}
\bibliography{dpm}

\newpage
\begin{table}[H]
\caption{Descriptive characteristics of the analysis sample at inclusion and over follow-up, MEMENTO cohort, France, 2011-2019 (N=2186)} \label{Tab1}
\centering
\begin{tabular}{llllll}
\toprule
\multicolumn{1}{c}{ } & \multicolumn{3}{c}{Inclusion in the study} & \multicolumn{2}{c}{Follow-up} \\
\cmidrule(l{3pt}r{3pt}){2-4} \cmidrule(l{3pt}r{3pt}){5-6}
Variable & Mean/n & SD/\% & N & Repeated measures & N$^*$ \\
\midrule
\addlinespace[0.3em]
\multicolumn{6}{l}{\textbf{Individuals characteristics}}\\
\hspace{1em}Age (years) & 70.9 & 8.7 & 2186 & \\
\hspace{1em}Female & 1349 & 61.7\% & 2186 & \\
\hspace{1em}$>$12 years education & 856 & 39.2\% & 2186 & \\
\hspace{1em}\textit{APOE4} carrier & 654 & 29.9\% & 2186 & \\
\addlinespace[0.6em]
\multicolumn{6}{l}{\hspace{1em}CDR sum of the boxes}\\
\hspace{2em}= 0 & 784 & 35.9\% & 2186 & \\
\hspace{2em}= 0.5 & 794 & 36.3\% & 2186 & \\
\hspace{2em}= 1 & 323 & 14.8\% & 2186 & \\
\hspace{2em}$>$ 1 & 285 & 13.0\% & 2186 & \\
\addlinespace[0.6em]
\multicolumn{6}{l}{\textbf{Markers of AD (CSF)}}\\
\hspace{1em}t-tau (pg/ml) & 376.0 & 264.8 & 341 & 1.3 & 395\\
\hspace{1em}p-tau (pg/ml) & 62.6 & 29.5 & 342 & 1.3 & 396\\
\hspace{1em}A$\beta_{42}$ (pg/ml) & 1096.2 & 411.2 & 342 & 1.3 & 396\\
\addlinespace[0.6em]
\multicolumn{6}{l}{\textbf{Brain imaging markers}}\\
\hspace{1em}WMH volume (mm3) & 9.9 & 13.2 & 1993 & 1.7 & 2056\\
\hspace{1em}Hippocampal volume (cm3) & 6.5 & 1.3 & 2047 & 1.7 & 2048\\
\hspace{1em}Entorhinal cortex (mm) & 3.3 & 0.4 & 2041 & 1.7 & 2057\\
\hspace{1em}Middle temporal cortex (mm) & 2.7 & 0.1 & 2047 & 1.7 & 2064\\
\hspace{1em}Fusiform cortex (mm) & 2.6 & 0.1 & 2047 & 1.7 & 2064\\
\hspace{1em}PET-FDG (SUVr) & 2.1 & 0.3 & 1236 & 1.5 & 1400\\
\addlinespace[0.6em]
\multicolumn{6}{l}{\textbf{Cognitive tests}}\\
\hspace{1em}Verbal Fluency (animals cited) & 28.3 & 8.8 & 2147 & 4.6 & 2164\\
\hspace{1em}TMT-A (good moves/sec) & 0.6 & 0.2 & 2165 & 4.6 & 2184\\
\hspace{1em}Free recalls FCSRT (sum score) & 26.1 & 8.3 & 2169 & 4.5 & 2176\\
\bottomrule
\end{tabular} 
\newline
SD = standard deviation, \textit{APOE4} = allele $\epsilon$4 of the apoliprotein E gene, CDR = clinical dementia rating, AD =Alzheimer's disease, CSF = cerebrospinal fluid, t-tau = total tau, p-tau = phosphorylated tau, A$\beta_{42}$ = isoform 42 of protein amyloid, WMH = white matter hyperintensities, PET = positron emission tomography, FDG = fluorodeoxyglucose, SUVr = standard uptake value ratio, TMT-A = trail making test A, FCSRT = free and cued selective reminding test\\
\end{table}

\begin{figure}[H]
    \includegraphics[width=1\textwidth]{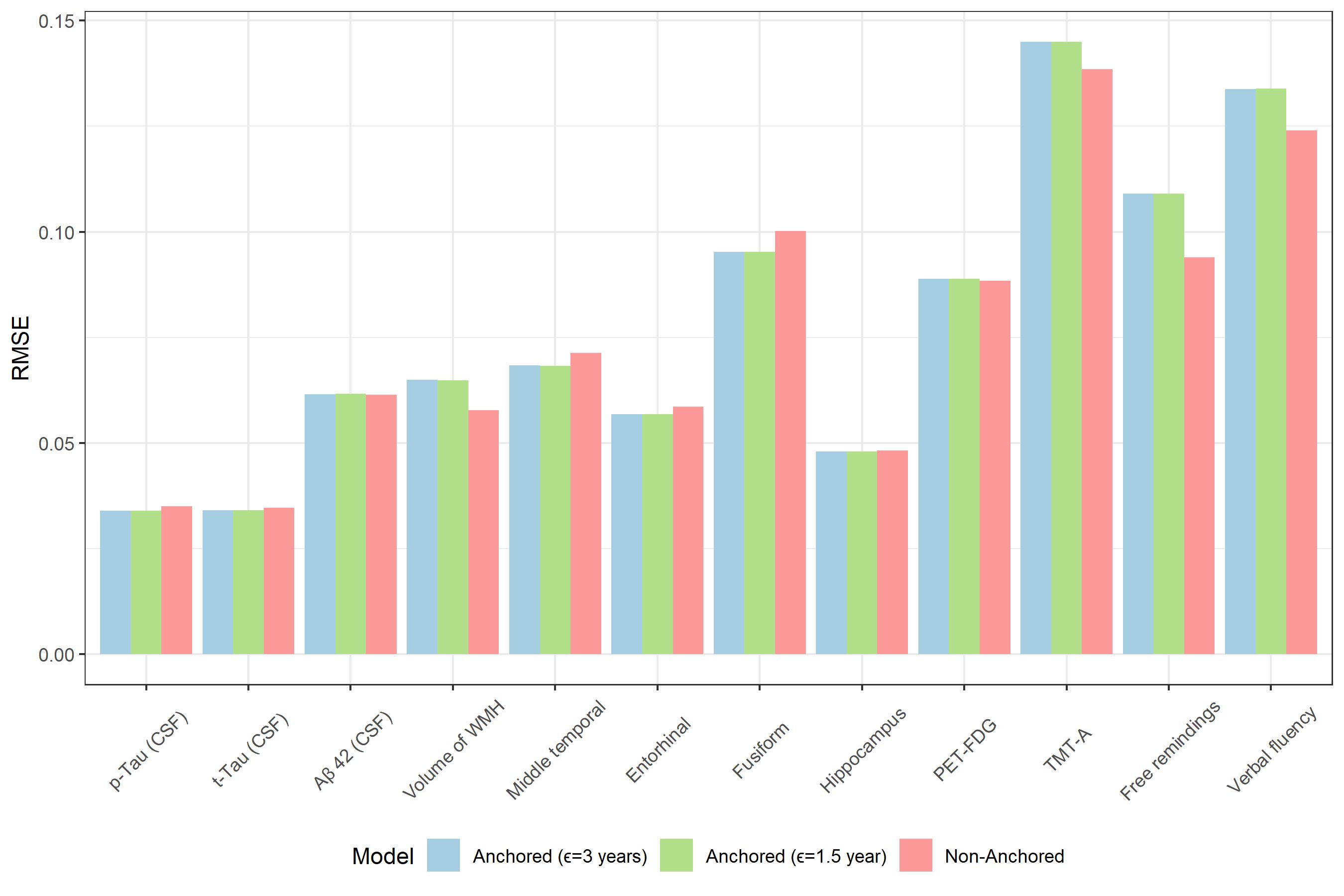}
    {e-Figure 1. Residual mean square error per marker for the main disease progression anchored model (DPAM) with constraint of 1.5 years around the observed clinical diagnosis, the DPAM with a weaker constraint of 3 years, and a non-anchored disease progression model, the MEMENTO Cohort, France, 2011-2019 (N=2186).}
    \label{fig:e-fig1}
\end{figure}

\begin{figure}[H]
    \includegraphics[width=1\textwidth]{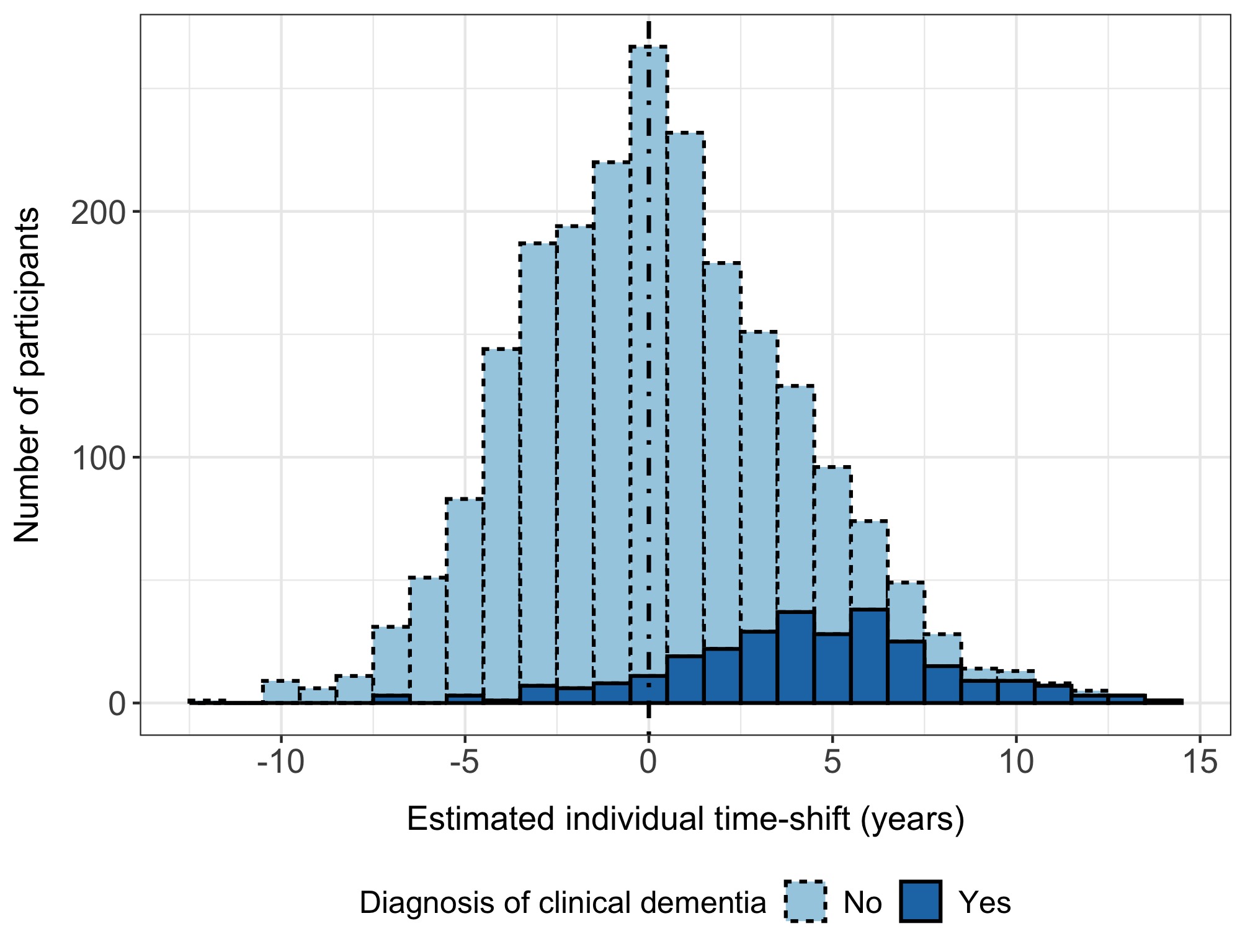}
    {e-Figure 2. Posterior distribution of the estimated individual time shifts from a non-anchored disease progression model, the MEMENTO Cohort, France, 2011-2019 (N=2186}
    \label{fig:e-fig2}
\end{figure}

\end{document}